\documentclass[aps,pra,preprint,groupedaddress]{revtex4-1}
\usepackage{amsmath} 
\usepackage{graphicx}
\usepackage{color} 
\usepackage{url}
\usepackage{listings}
\begin{document}


\title{Solve the Master Equation by Python-An Introduction to the
Python Computing Environment}


\author{Wei~Fan} 
\author{Yan~Xu}
\author{Bing~Chen}
\author{Qianqian~Ye}
\affiliation{College of Science, Shandong University of Science and
Technology, Qingdao 266510, China}

\date{\today}

\begin{abstract}
A brief  introduction to the Python computing
environment is given. By solving the master equation encountered in
quantum transport, we  give an example of how to solve the ODE
problems in Python. The ODE solvers used are the ZVODE routine in
Scipy and the bsimp solver in GSL. For the former, the equation can be
in its complex-valued form, while for the latter, it has to be
rewritten to a real-valued form. The focus is on the detailed workflow
of the implementation process, rather than on the syntax of the python
language, with the hope to help  readers  simulate their own models in
Python.  
\end{abstract}


\maketitle

\section{Introduction}

Python is a general purpose, high-level programming language. It is
well documented and easy to learn. With its concise and high-readable
code, it greatly improves the efficiency of code development and code
reuse, so we can save our time by simulating our model in Python.

In computation, Python provides us with an easy-to-use environment
which saves both the code development time and the code execution
time.  Though it is an interpreted language, which means its execution
speed is lower compared with the compiled language like C and Fortran,
its performance has been greatly improved by a set of specific
packages designed for numerical computation. This is done by gluing
well behaved numerical libraries written in C/C++ and Fortran and
allowing users to use them in Python code, thus we can write code
efficiently in Python and get an execution speed like C and Fortran.
Those packages includes Numpy~\cite{scipy,computingScience},
Scipy~\cite{scipy,computingScience}, Pygsl~\cite{pygsl},
Cython~\cite{cbehnel2010cython} and various others packages.  Numpy is
based on the compiled LAPACK libraries that is standard for linear
algebra computations. It provides the  data structure 'array' and fast
operations on arrays, such as linear algebra, Fourier transform and
random number generation, which  make it easier and faster to handle
matrix related problems. As the computation is essentially executed in
LAPACK, it can run almost as fast as in the C code.  Based on Numpy,
Scipy provides many modules to perform the common tasks in science and
industry, such as FFT, sparse matrix, statistics, signal processing
and ODE solvers. The functionality of Numpy and Scipy is similar to
Matlab,  but they are developed to make scientific computing a natural
part of Python, rather than be a copy of Matlab.  Pygsl is a python
interface to the GNU Scientific Library (GSL)~\cite{gslRef}, which is
an open source C library for numerical computations in science. The
models can be expresed in Numpy arrays and then by Pygsl we can use
the functions in GSL directly as if they are Python functions, which
greatly simplifies the usage of GSL functions.  Cython is based on
Python, but it allows for static C type declarations and direct
calling of C or C++ functions, which combines the high productivity of
Python with the execution speed of C~\cite{SciPyProceedings_4}. Cython
is used in the development of the powerful computing software
sage~\cite{sage} (or sagemath).  Besides computation, there are many
powerful visualization tools in Python, such as Visual~\cite{vpython}
for animation and Matplotlib~\cite{matplot} for Matlab-like plot.
        Other plotting tools, such as Mayavi~\cite{mayavi2} for 3D
        visulization and gnuplot~\cite{gnuplot} for scientific plot,
        can also be used in Python directly.

Even though there is very few cases where no package exists for
performance improvement, we still recommend Python as it could save
lots of our time in code development, while using the compiled
languages can take lots of human time and would be exhausting. On the
other hand, Python allows us to write the time consuming part of the
algorithm, usually long loops, directly in compiled languages like
C/C++ and Fortran, so we can get an optimal balance between human time
and machine time. This can be done by lots of powerful wrappers, such
as F2PY~\cite{f2py}, which can automatically wrap Fortran codes and
make it callable from Python, and Weave~\cite{Weave}, which makes it
possible to written C and C++  codes directly in Python codes.

In this paper, we will use Python to solve the master equation in
reference~\cite{PhysRevB.70.235317}, which is an ordinary differential
equation (ODE). This is an introduction to the Python computing
environment and a detailed example of solving the ODE problems in
Python. We will use three solvers, the ZVODE routine in Scipy, the
bsimp solver and the rkf45 solver of GSL to solve it. We will also
give an example of using Cython to optimize the code to get faster
execution speed.  The emphasis is layed on the workflow of how to
solve the ODE problems in Python rather  than on the syntax of this
language. One can turn to the
books~\cite{Fangohr,landau2008,kiusalaas2010} for a detailed
tutorial of computational physics in Python.

\section{The Master Equation}

The master equation to be solved is
\begin{equation}\label{masterEquation} 
\dot{\rho} = - i \left[
\mathcal{H}, \rho \right] - \Gamma \left[ \rho - diag(\rho) \right],
\end{equation}
where the Hamiltonian $\mathcal{H}=\left(
\begin{array}{ccc}0 & -\Omega_{12} & 0\\ -\Omega_{12} & 0 &
-\Omega_{23}\\ 0 & -\Omega_{23} & 0 \end{array}\right) $ describes the
coherent tunneling via adiabatic passage (CTAP) scheme, $\Gamma$ is
the $T_2$ dephasing rate and $\Omega_{12}$ and $\Omega_{23}$ are the
tunneling rate between the corresponding quantum dots 
\begin{eqnarray}
\Omega_{12}\left( t\right) & = & \Omega^{max} \exp\left[ - \left( t -
\frac{t_{max} + \sigma}{2} \right) ^2 / \left(2 \sigma^2 \right)
\right]\nonumber\\ \Omega_{23}\left(t\right) & = &
\Omega^{max}\exp\left[ - \left( t - \frac{t_{max} -
\sigma}{2}\right)^2 / \left( 2 \sigma^2 \right) \right].\nonumber
\end{eqnarray}
This is an evolution equation of the density matrix of
the system, which describes the evolution of the electron in the dots.
The master equation is often used when the decoherence effect are
considered due to the interaction between the system and its
environment. The diagonal terms of the density matrix describe the
population of the electron in the corresponding dots, while the
off-diagonal terms describe the correlation between corresponding
dots. For the $T_2$ dephasing term, the off-diagonal terms describe
the decoherence of the electron, which corresponds to the loose of its
quantum nature. The neglected diagonal terms in the dephasing terms
describe the loss of particles in corresponding dots due to the
interaction with the environment, which is not important in the CTAP
scheme. 

To judge the efficiency of the transporting system, this equation
needs to be solved.  We would first solve the case without the
dephasing term using the solver ZVODE and bsimp, and only consider the
case of counter-intuitive pulses, that is, the result shown in
Figure.3(b) in reference~\cite{PhysRevB.70.235317}. After that, we
would solve the case with the dephasing term included using the rkf45
solver and give the pseudo-color plot correspoding Figure.4 in
reference~\cite{PhysRevB.70.235317}.  The parameters are chosen  as
the same as there with $\sigma = t_{max} / 8$.  Reducing $t$ by
$\pi/\Omega^{max}$, that is,  $t=t/\left(\pi/\Omega^{max}\right)$ and
$t_m=t_{max}/\left(\pi/\Omega^{max}\right)$, and then rescale $t$ to 1
by $t=t/t_{m}$, the equation is reduced to the form
\begin{equation}
\label{firstForm} 
\dot{\rho} = -i \pi t_m
[\mathcal{H}, \rho] - \Gamma \pi t_m [\rho - diag(\rho)],
\end{equation}
where $\Gamma = \Gamma / \Omega^{max}$ and the coupling
pulses are reduced to 
\begin{eqnarray}
\Omega_{12}\left( t\right) & =
&  \exp\left[ - 32 \left( t - \frac{9}{16} \right) ^2
\right]\nonumber\\ \Omega_{23}\left(t\right) & = &  \exp\left[ - 32
\left( t  - \frac{7}{16}\right)^2  \right].  
\end{eqnarray} This is
the equation we need to solve in Python.

\section{The Solvers}

Scipy provides two ODE solvers 'odeint' and 'ode' in its 'integrate'
module, with the former using the 'lsoda' of the Fortran library
odepack and the latter using the VODE ( for real-valued equations) and
the ZVODE (for complex-valued equations) routines.  The ZVODE routine
provides the implicit Adams method for non-stiff problems and a method
based on the backward odifferentiation formulas (BDF) for stiff
problems.  Stiff equation is that includes some terms that can lead to
rapid vibration in the solution and it requires the step size to be
taken extremely small if using non-stiff solvers. This would happen
when the variables are changing on two vastly different scales. If you
do not know whether you problem is stiff or not, using the stiff
solvers would be safe, although it would cost more time.  Here
we will use the ZVODE routine with the BDF method. The CTAP
scheme is non-stiff, but we use stiff solvers here just to
give examples of how to use them. In the last part when we
solve the equation with the dephasing effect, we would use the
non-stiff solver and optimize it by Cython, as there is a hard
requirement on computation time there. As the solver only
accepts a set of equations that are in a vector form, we have
to rewrite equation \eqref{firstForm} from the matrix form to
a one-dimensional array form. This can be done via the Numpy
array and related operations.  The equation finally should be
in the form \begin{equation}\label{odeComplex} \frac{dy}{dt}
= - i \pi t_m f, \end{equation} where $y$ is the column
version of $\rho$ and $f$ is the derivative of $y$ with time. 

\begin{figure} 
\begin{center} 
\includegraphics{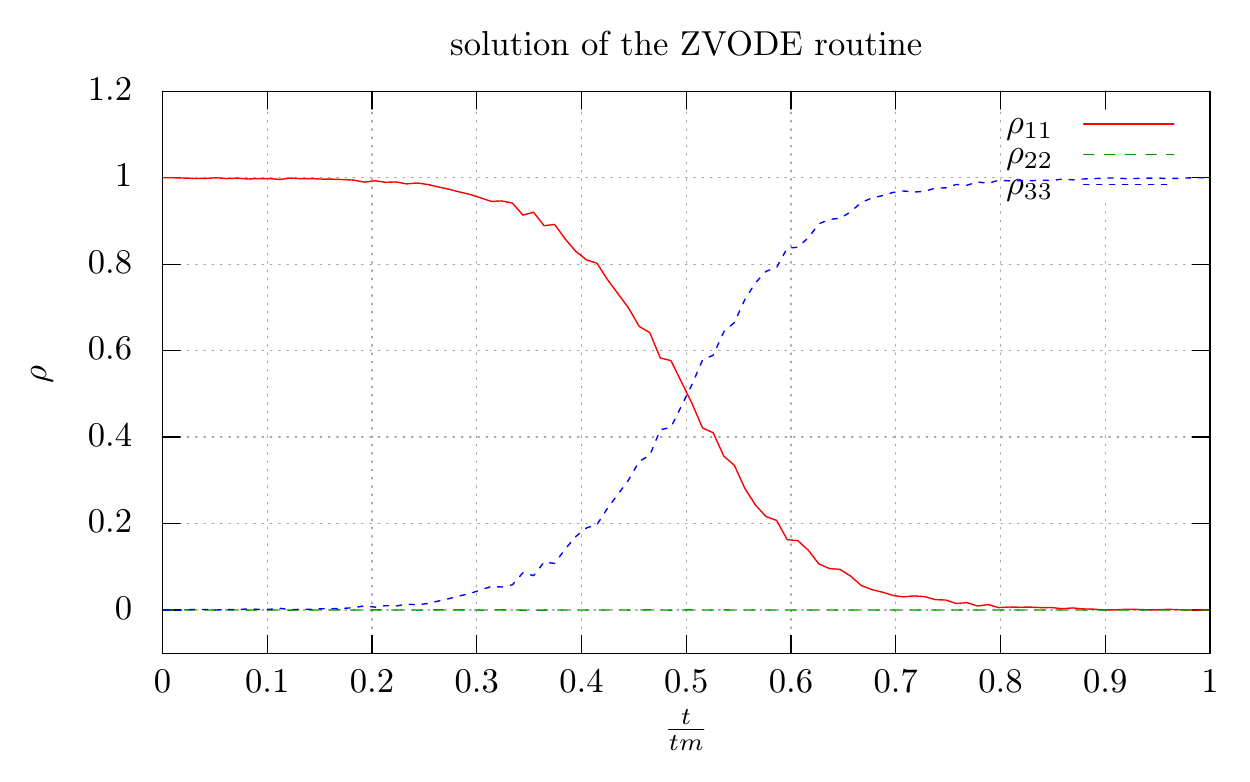}
\caption{Solution got via the ZVODE routine for the counter-intuitive
pulse without dephasing. The particle evolves from the first dot to
the third dot without any population on the second dot. This is  the
result of the CTAP scheme, as explained in
reference~\cite{PhysRevB.70.235317}. Here $t_{m}$ has been set to 8000
to avoid vibrations of the  lines.} 
\label{zvodeFig} 
\end{center}
\end{figure}

Listing \ref{zvode.py} impliments the solution of equation
\eqref{odeComplex}. We first import it by 'from Scipy.integrate
import ode', which imports the solver ode from Scipy's integrate
module. Then we define equation \eqref{odeComplex}  and its Jacobian
using  the data structure array provided by Numpy,  which is
straightforward in describing   matrix related problems . The
integrator of the solver should be set to use the ZVODE routine with
the BDF method. After setting the step size and the precision
anticipated, we  can call this solver to forward the calculation step
by step until it reaches the final value. We can plot the data
directly in Python using Matplotlib or gnuplot~\cite{gnuplot.py}.
Here we output the data to a file and plot it in gnuplot. The source
file of gnuplot would also be given in listing \ref{plotZvode.plt}
and the result is shown in Figure.~\ref{zvodeFig}.
\lstset{language=Python, caption={This program solves the CTAP 
regime without the dephasing term using the zvode routine of Scipy.}, 
label=zvode.py}
\begin{lstlisting}
#! /usr/bin/env python2.7
#First import the modulus used in this program. Numpy is so often used that perhaps every 
#program should import it first. The ode solver is in Scipy's integrate modulus.

import numpy                             
from scipy.integrate import ode  

#Define function dy/dt = f. We have to write the elements of f out by hand, as 
#this solver requires the Jacobian, whose elements must be calculated by hand. 
#For the rkf45 solver used below, which does not require the Jacobian, we can express it by 
#Numpy array operations without calculating its elements by hand.
def userSupply(t, y, tm):
    j1=numpy.exp(-32.*(t-9./16)**2)   
    j2=numpy.exp(-32.*(t-7./16)**2)
    #complex number is represented by j in Python. 5j means 5*i and 1j means i.
    return -1j*numpy.pi*tm*numpy.array(
        [j1*y[1]-j1*y[3], j1*y[0]+j2*y[2]-j1*y[4], j2*y[1]-j1*y[5],
         j1*y[4]-j1*y[0]-j2*y[6], j1*y[3]+j2*y[5]-j1*y[1]-j2*y[7], 
         j2*y[4]-j1*y[2]-j2*y[8], j1*y[7]-j2*y[3], 
         j1*y[6]+j2*y[8]-j2*y[4], j2*y[7]-j2*y[5]])

#The jacobian df/dy of the ODE equations.
def jac(t, y, tm):                     
    j1=numpy.exp(-32.*(t-9./16)**2)
    j2=numpy.exp(-32.*(t-7./16)**2)
    return -1j*numpy.pi*tm*numpy.array([[0, j1, 0, -j1, 0, 0, 0, 0, 0],
                                       [j1, 0, j2, 0, -j1, 0, 0, 0, 0],
                                       [0, j2, 0, 0, 0, -j1, 0, 0, 0],
                                       [-j1, 0, 0, 0, j1, 0, -j2, 0, 0],
                                       [0, -j1, 0, j1, 0, j2, 0, -j2, 0],
                                       [0, 0, -j1, 0, j2, 0, 0, 0, -j2],
                                       [0, 0, 0, -j2, 0, 0, 0, j1, 0],
                                       [0, 0, 0, 0, -j2, 0, j1, 0, j2],
                                       [0, 0, 0, 0, 0, -j2, 0, j2, 0]])
#initial time and value.
t0 = 0
y0 = numpy.array([1, 0, 0, 0, 0, 0, 0, 0, 0], dtype=numpy.complex128) 
#t_max and the step size.
tm = 8000
H = 1.e-3
#Call the solver ode with the function 'userSupply' and the Jacobian 'jac', and
#define the integrator 'zvode' with the 'bdf' method and set tolerance 'rtol'.
r = ode(userSupply, jac).set_integrator(
    'zvode', method='bdf', with_jacobian=True, rtol=1.e-6, order=5) 
#Prepare the intial values and the parameters t_max.
r.set_initial_value(y0, t0).set_f_params(tm).set_jac_params(tm)

#define the file to write data to.
output1=open('dataCounBdf','w')    
#Forward the integrator in a loop and output the data to the file.
while r.successful() and r.t < 1.:
    if ( 1. - r.t) < H: H = 1. -r.t
    r.integrate(r.t + H)
    output1.write("%f %f %f %f\n" % (r.t, r.y[0], r.y[4], r.y[8]))

output1.close() #close the file.

\end{lstlisting}

\lstset{language=Gnuplot,caption={The program plots
the evolution result.},label=plotZvode.plt}
\begin{lstlisting}
set terminal tikz standalone color size 5in,3in
set out 'counterBdf.tex'
set xlabel '$\frac{t}{tm}$'
set ylabel '$\rho$'
set yrange [-0.1:1.2]
set grid
set title 'solution of the ZVODE routine'
plot 'dataCounBdf' using ($1):($2) smooth csplines title '$\rho_{11}$', 'dataCounBdf' using ($1):($3) smooth csplines title '$\rho_{22}$', 'dataCounBdf' using ($1):($4) smooth csplines title '$\rho_{33}$'
\end{lstlisting}

\begin{figure} 
\begin{center} 
\includegraphics{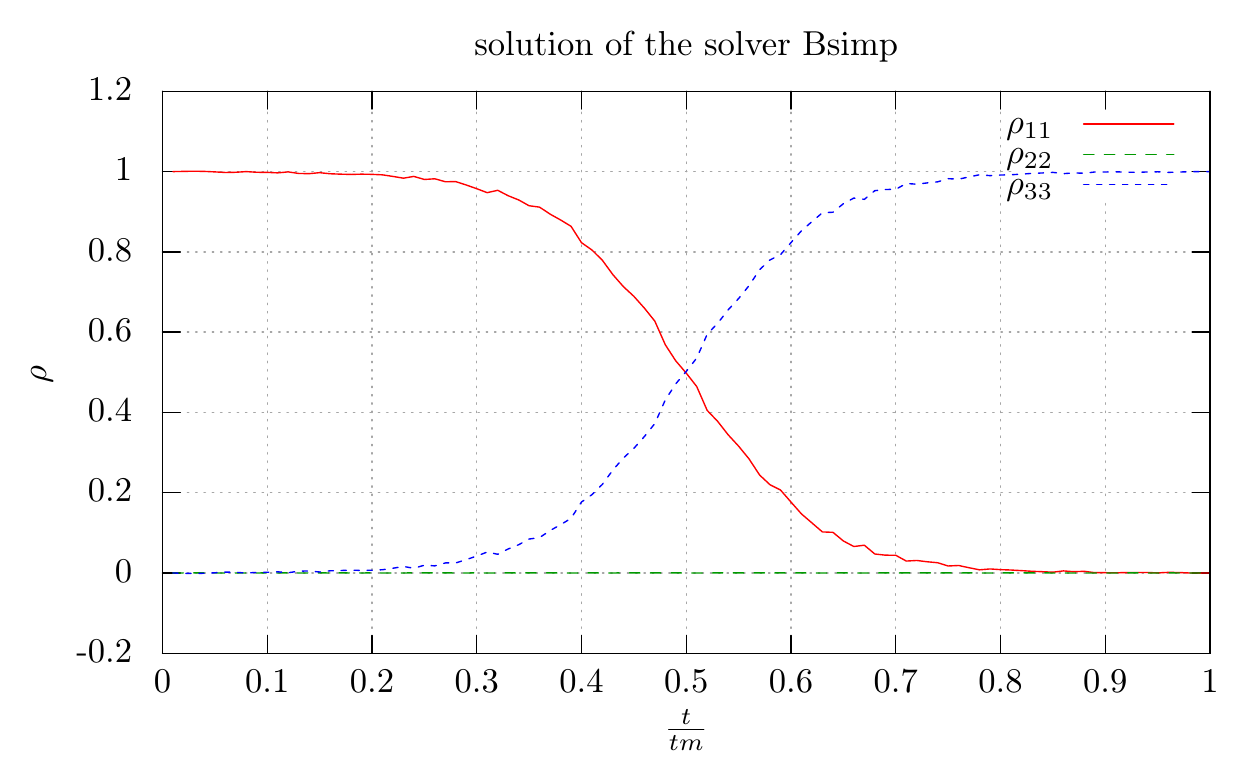}
\caption{Solution got via the bsimp solver of GSL for the
counter-intuitive pulse without dephasing. It's the same as
Figure.~\ref{zvodeFig}. $t_{m}$ also equals to 8000.} 
\label{bsimpFig}
\end{center} 
\end{figure}

GSL provides many ODE solvers, using the implicit Bulirsch-Stoer
method, the Gear method and various Runge-Kutta methods. They can be
used in Python via the Pygsl module, which makes  their usage
straightforward, just like using the solvers in Scipy, and you can
read its document by typing 'help('pygsl.odeiv')' in Python. Here the
solver bsimp is used, which is given in listing \ref{bsimp.py}.  The
bsimp solver impliments the implicit Bulirsch-Stoer method of Bader
and Deuflhard.  For smooth functions, the Bulirsch-Stoer method is the
best way to achieve both high-accuracy solutions and computational
efficiency~\cite{press1994numerical}, with the implicit Bulirsch-Stoer
method designed for stiff problems.  As above, the Pygsl module should
to be imported first. Then is the definition of the equation and its
Jacobian. It should be noticed that the solvers in GSL only accept
real-valued equations, so $y$'s  real and imaginary part in
equation~\eqref{odeComplex} have to be separated and form a new set of
coupled  equations with 18  elements, with $y[2 i]$ and $y[2 i + 1]$
the real and the imaginary part  of the original$y[i]$.   Their
preparation is a little different from solvers in Scipy, which can be
learned from its documentation. It only needs inputs of the initial
step size and the errors tolerated, as the following  step sizes are
adjusted automatically to optimize its speed and precision.  The
gnuplot source code is as the code \ref{plotZvode.plt} except with a
different input data name and output file name, so we do not put it
here.  The result is given in Figure.~\ref{bsimpFig}.
\lstset{language=Python,caption={This program solves the CTAP regime
without the dephasing using the bsimp solver of GSL.},label=bsimp.py}
\begin{lstlisting}
#! /usr/bin/env python2.7
#First import the modulus used in this program. 
#The solver is in pygsl's odeiv module.
import numpy
from pygsl import odeiv

#User supplied function dy/dt = f. This is more complex compared with the above code.
#As we have to seperate the real and imaginary parts from the original equations.
def userSupply(t, y, tm):
    j2 = numpy.exp(-32.*(t-9./16)**2)
    j1 = numpy.exp(-32.*(t-7./16)**2)
    return numpy.pi*tm*numpy.array(
        [j1*(y[3]-y[7]), j1*(y[6]-y[2]), j1*(y[1]-y[9])+j2*y[5],
         j1*(y[8]-y[0])-j2*y[4], j2*y[3]-j1*y[11], j1*y[10]-j2*y[2], 
         j1*(y[9]-y[1])-j2*y[13], j1*(y[0]-y[8])+j2*y[12],
         j1*(y[7]-y[3])+j2*(y[11]-y[15]), j1*(y[2]-y[6])+j2*(y[14]-y[10]),
         j2*(y[9]-y[17])-j1*y[5], j2*(y[16]-y[8])+j1*y[4], 
         j1*y[15]-j2*y[7], j2*y[6]-j1*y[14], j1*y[13]+j2*(y[17]-y[9]), 
         -j1*y[12]+j2*(y[8]-y[16]), j2*(y[15]-y[11]), j2*(y[10]-y[14])])

#the Jacobian df/dy and df/dt. The bsimp solver requires both df/dy and df/dt.
def jac(t, y, tm):
    j2 = numpy.exp(-32.*(t-9./16)**2)
    j1 = numpy.exp(-32.*(t-7./16)**2)
    dj2dt = -64*(t-9./16)*j2
    dj1dt = -64*(t-7./16)*j1
    dfdy = numpy.zeros([18,18])
    dfdy[0, 3] = j1; dfdy[0, 7] = -j1;
    dfdy[1, 2] = -j1; dfdy[1, 6] = j1;
    dfdy[2, 1] = j1; dfdy[2, 5] = j2; dfdy[2, 9] = -j1
    dfdy[3, 0] = -j1; dfdy[3, 4] = -j2; dfdy[3, 8] = j1
    dfdy[4, 3] = j2; dfdy[4, 11] = -j1
    dfdy[5, 2] = -j2; dfdy[5,10] = j1
    dfdy[6, 1] = -j1; dfdy[6, 9] = j1; dfdy[6, 13] = -j2
    dfdy[7, 0] = j1; dfdy[7, 8] = -j1; dfdy[7, 12] =j2
    dfdy[8, 3] = -j1; dfdy[8, 7] = j1; dfdy[8, 11] = j2; dfdy[8, 15] = -j2
    dfdy[9, 2] = j1; dfdy[9, 6] = -j1; dfdy[9, 10] = -j2; dfdy[9, 14] = j2
    dfdy[10, 5] = -j1; dfdy[10, 9] = j2; dfdy[10, 17] = -j2
    dfdy[11, 4] = j1; dfdy[11, 8] = -j2; dfdy[11, 16] = j2
    dfdy[12, 7] = -j2; dfdy[12, 15] = j1
    dfdy[13, 6] = j2; dfdy[13, 14] = -j1
    dfdy[14, 9] = -j2; dfdy[14, 13] = j1; dfdy[14, 17] = j2
    dfdy[15, 8] = j2; dfdy[15, 12] = -j1; dfdy[15, 16] = -j2
    dfdy[16, 11] = -j2; dfdy[16, 15] = j2
    dfdy[17, 10] = j2; dfdy[17, 14] = -j2
    dfdt = numpy.array(
        [dj1dt*(y[3]-y[7]), dj1dt*(y[6]-y[2]), dj1dt*(y[1]-y[9])+dj2dt*y[5],
         dj1dt*(y[8]-y[0])-dj2dt*y[4], dj2dt*y[3]-dj1dt*y[11], dj1dt*y[10]-dj2dt*y[2], 
         dj1dt*(y[9]-y[1])-dj2dt*y[13], dj1dt*(y[0]-y[8])+dj2dt*y[12],
         dj1dt*(y[7]-y[3])+dj2dt*(y[11]-y[15]), dj1dt*(y[2]-y[6])+dj2dt*(y[14]-y[10]),
         dj2dt*(y[9]-y[17])-dj1dt*y[5], dj2dt*(y[16]-y[8])+dj1dt*y[4], 
         dj1dt*y[15]-dj2dt*y[7], dj2dt*y[6]-dj1dt*y[14], dj1dt*y[13]+dj2dt*(y[17]-y[9]), 
         -dj1dt*y[12]+dj2dt*(y[8]-y[16]), dj2dt*(y[15]-y[11]), dj2dt*(y[10]-y[14])])
    return numpy.pi*tm*dfdy, numpy.pi*tm* dfdt

#Initial values.
dimension = 18
t = 0 
y = numpy.array([1, 0, 0, 0, 0, 0, 0, 0, 0, 0, 0, 0, 0, 0, 0, 0, 0, 0])
tm = 50

#the stepping function advances the solution from t_i to t_i+1 .
#step_bsimp is the bsimp solver we use.
step = odeiv.step_bsimp(dimension,userSupply,jac,args=tm)
#the control function  optimizes the step size with the input tolerated errors.
control = odeiv.control_yp_new(step,1.e-6,1.e-6)
#Based on the stepping and the control funtion, the evolve
#function advances the solution in a given interval.
evolve = odeiv.evolve(step, control,dimension)

#The file to store data.
output1=open('dataInBsipmp','w')

#initial step size. It can optimize step size under given errors. Only need to input an initial step size.
h = 1.e-2
while t < 1.:
    if (1. - t) < h: h=1. - t
    t, h, y = evolve.apply(t, 1, h, y)
    output1.write("%f %f %f %f\n" % (t, numpy.sqrt(y[0]**2+y[1]**2), numpy.sqrt(y[8]**2+y[9]**2), numpy.sqrt(y[16]**2+y[17]**2)))
    
#close file.
output1.close()

\end{lstlisting}

Now we add the dephasing term and solve the equation
\eqref{firstForm}. This is done by the rkf45 solver of GSL, which uses
the Runge-Kutta-Fehlberg method with adaptive stepsize control using
the 5th order error estimate. The rkf45 method is the most general
method and applies to almost all initial value ODE problems. The
solution process gets greatly simplified using this solver, as it does
not require the Jacobian of the system, the calculation of which takes
most of our time. The Jacobian needs to do partial differentiation on
the funtion $f$ of equation \eqref{odeComplex}, which needs to be done
by hand. Without it, we can use Numpy array operations to write the
equation \eqref{firstForm} to the one-dimensional array form
\eqref{odeComplex} and do not need to calculate by hand at all, thus
greatly saves our time. This can be done by substituting $\rho=\rho_R
+ i \rho_C$ into equation \eqref{firstForm} and seperating the
derivative of the $\rho_R$ and $\rho_C$. After rescaling, we would get
the following equation 
\begin{eqnarray}
\frac{d\rho_R}{dt}=[H,\rho_C]-\Gamma[\rho_R-diag(\rho_R)]\nonumber\\
\frac{d\rho_C}{dt}=-[H,\rho_R]-\Gamma[\rho_C-diag(\rho_C)].
\end{eqnarray} 
This form can be expressed to the one-dimensional form
easily in Numpy, as the listing \ref{dephase.py} shows, which solves
the equation with dephasing included.

\begin{figure}
\begin{center} 
\includegraphics[height=3.40in]{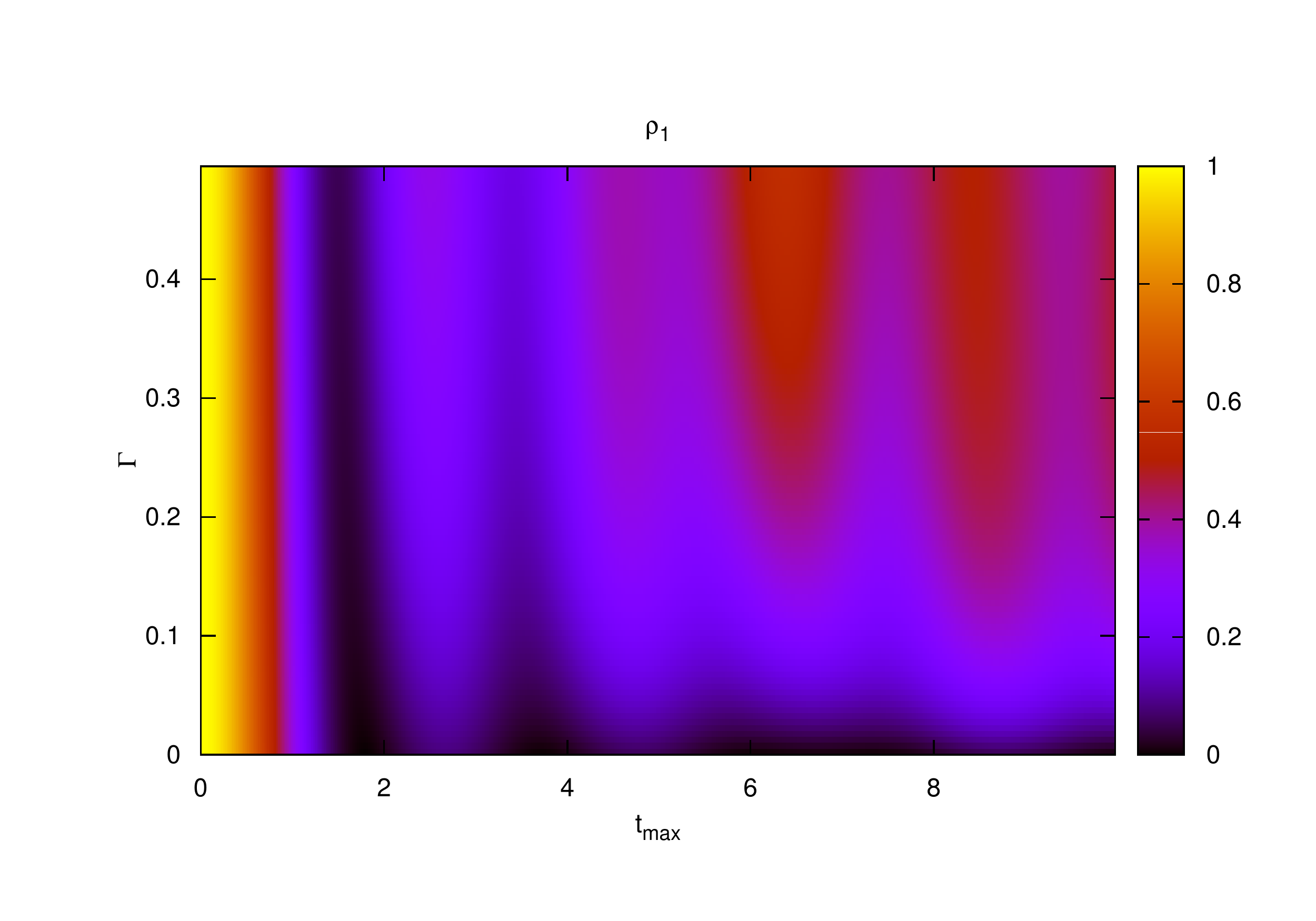}\\
\vspace{-50pt}
\includegraphics[height=3.40in]{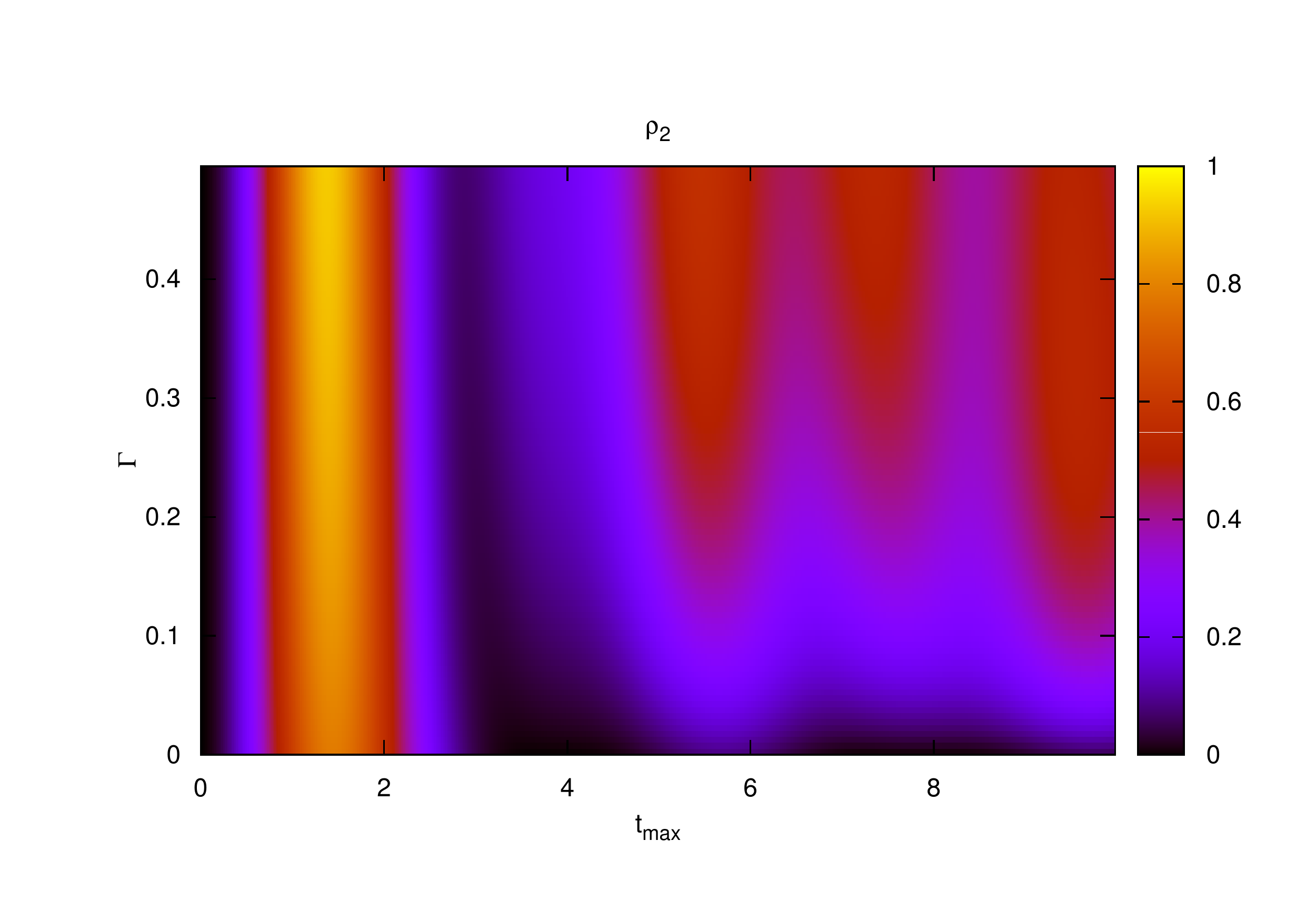}\\
\vspace{-50pt}
\includegraphics[height=3.40in]{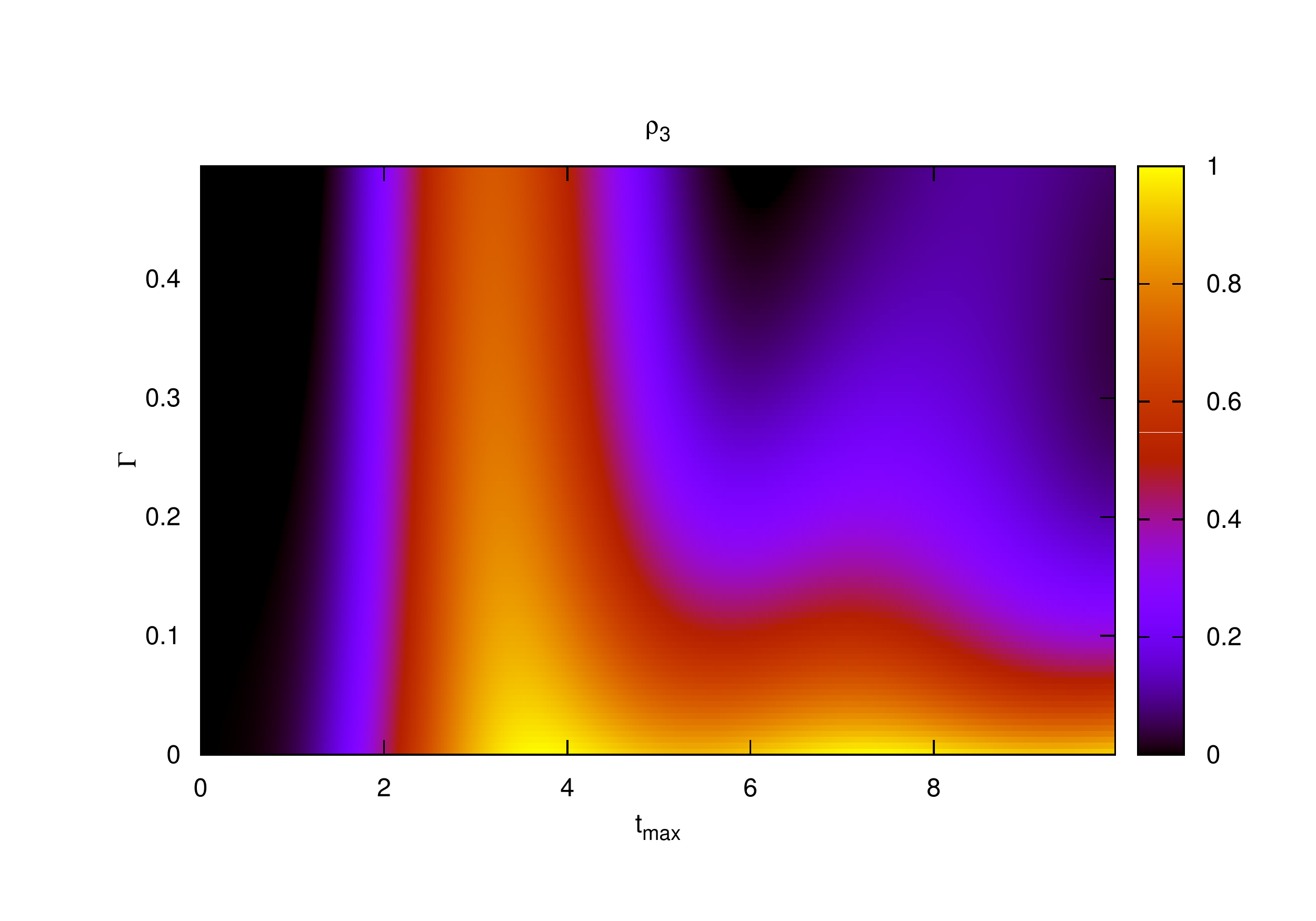}
\vspace{-30pt}
\caption{The pseudo-color
plot of the master equation including the dephasing. See reference \cite{PhysRevB.70.235317}. }
\label{dephaseFig} 
\end{center} 
\end{figure}

To get the pseudo-color plot of Figure. 4 of reference
\cite{PhysRevB.70.235317}.  we need to solve the master equation
\eqref{firstForm} many times with different values of $\Gamma$ and
$t_m$ each time. Here we solve it for $100x500$ times with $100$
different $\Gamma$ and $500$ different $t_m$. This is done in a loop
and would cost huge computation time. We would give an Cython code to
optimize the Python code, which is also an example of how to use
Cython.  The optimzed code with Cython is given in listing 
\ref{dephaseCython.pyx} in the next section. The gnuplot source code
is given in listing \ref{plotDephase.plt} and the pseudo-color plot is
shown in Figure. \ref{dephaseFig}.
\lstset{language=Python,caption={This program solves the CTAP regime
with the dephasing included using the solver rkf45 of
GSL.},label=dephase.py} 
\begin{lstlisting}
#! /usr/bin/env python2.7
#First import the modulus used in this program. 
#The solver rkf45 is in pygsl's odeiv module.
import numpy
from pygsl import odeiv

#Abstract class used for declaration.
class Function:
        def evaluate(self,t,y,lis):
                tm,x=lis
                return 0
#User supplied function dy/dt = f. We can use the Numpy array and related operations to define it, including array reshape and concatenate, which simplifys the process.
class userSupply(Function):
        def evaluate(self,t,y,lis):
                tm,x=lis
                j1 = numpy.exp(-32.*(t-9./16)**2)
                j2 = numpy.exp(-32.*(t-7./16)**2)
                H=numpy.array([[0,j1,0],[j1,0,j2],[0,j2,0]])
                rhoR=y[0:9].reshape((3,3))
                rhoC=y[9:18].reshape((3,3))
                dRdt=numpy.dot(H,rhoC)-numpy.dot(rhoC,H)-x*(rhoR-numpy.diag(rhoR))
                dCdt=numpy.dot(rhoR,H)-numpy.dot(H,rhoR)-x*(rhoC-numpy.diag(rhoC))
                return numpy.pi*tm*numpy.concatenate((dRdt.reshape((9)),dCdt.reshape((9))))

#compute_dephase impliments the loop within which the equation is solved with different parameters each time.
def compute_dephase(f,N1,N2):
        output1=open('dataDephase1','w')
        output2=open('dataDephase2','w')
        output3=open('dataDephase3','w')
        dimension=18
        for x in numpy.linspace(0.,0.5,N1):
                for tm in numpy.linspace(0,10,N2):
                        step = odeiv.step_rkf45(dimension,f.evaluate,args=(tm, x))
                        control = odeiv.control_yp_new(step,1e-7,1e-7)
                        evolve = odeiv.evolve(step, control,dimension)
                        y=numpy.zeros((dimension))
                        y[0]=1.0
                        t=0.;h = 1.e-4
                        while t<1.:
                            if (1.-t)<h: h=1.-t
                            t, h, y = evolve.apply(t,1.,h,y)
                        output1.write(" %f " % y[0])
                        output2.write(" %f " % y[4])
                        output3.write(" %f " % y[8])
                output1.write("\n")
                output2.write("\n")
                output3.write("\n")
        output1.close()
        output2.close()
        output3.close()

compute_dephase(userSupply(),2,3)
\end{lstlisting}

\lstset{language=Gnuplot,caption={The program plots
the pseudo-color plot for the master equation with
dephasing.},label=plotDephase.plt} 
\begin{lstlisting}
set terminal post color enhanced 
set out 'dephase1.ps'
set xlabel 't_{max}'
set ylabel '{/Symbol G}'
set xtics ("0" 0, "2" 100,"4" 200,"6" 300,"8" 400, "10" 500)
set ytics ("0" 0, "0.1" 20,"0.2" 40,"0.3" 60,"0.4" 80, "0.5" 100)
set cbrange [0:1]
set pm3d map
set title '{/Symbol r}_{1}'
splot 'dataDephase1' matrix
exit
\end{lstlisting}

\section{Cython}

Cython allows C type declarations both for variables and functions,
and you can use C functions from C libraries in Cython. It can improve
the Numpy array operations, especially array iteration, to the speed
of C array operations, so we can write code efficiently using Numpy
array while get the speed of C. To use Cython, there are generally
three stages. Firstly we typedefine the variables to C static types
via the key word 'cdef', which removes the dynamical nature of those
variables and greatly improves the operation speed involving them. If
their are Numpy arrays, we should also typedefine their type, which is
shown in the listing \ref{dephaseCython.pyx}. There are a static type
and a dynamical type, both of which should be typedefined. After this
first stage, the program would get a faster speed compared with the
original one, especially when the iterator in the long loops are
typedefined. The second stage is to typedefine the Python functions to
return static type value via the  key word 'cdef type-used
function(tpye-used variable)', where the type-used are appropriate
types. This can greatly improve the speed of function calls, but now
this function can only be called in Cython. If using the key word
'cpdef', then the function can be called in Python. The function type
defination can improve the behavior of the total code greatly. The
last stage is to turn off some regular checks for some functions in
the code, especially for inline functions. Well optimized Cython code
can get an speed as fast as C and Fortran, as can be seen in the
examples provided by the Cython website. For detailed documentation,
turn to the reference ~\cite{cbehnel2010cython,SciPyProceedings_4} and
references there.  The listing \ref{dephaseCython.pyx} below only
serves as an example of how to use Cython. The list
\ref{profileDephase} shows the results of profiling the Python code
and the Cython code. The cpu is '2 Intel(R) Core(TM)2 Duo CPU T7500
@2.20GHz'. We can see that the Cython code saves about one third time
compared with the Pythond code. The speed got is not very much
compared with hundreds of times in examples provided by Cython
documentation. There are two reasons for this. The first is that this
is not optimizing a pure Python code. We have used many well behaved
packages in the Python code, which is very fast already. The second is
that our Cython code is not perfect. Many modifications can be done to
optimize it further.   

\lstset{language=Python,caption={The Cythond code to optimize the
Python one to get speed up.},label=dephaseCython.pyx}
\begin{lstlisting}
#! /usr/bin/env python2.7
#First import the modulus used in this program. 
#The solver rkf45 is in pygsl's odeiv module.

import numpy as np
cimport numpy as np
from pygsl import odeiv
cimport cython

cdef extern from "math.h":
        double exp(double)
@cython.profile(False)
cdef inline double EXP(double x):
        return exp(x)

DTYPE=np.float64
ctypedef np.float64_t DTYPE_t

#Abstract class used for declaration.
cdef class Function:
        cpdef np.ndarray[DTYPE_t,ndim=1] evaluate(self,double t,np.ndarray[DTYPE_t,ndim=1] y, np.ndarray[DTYPE_t,ndim=1] lis):
                cdef tm,x
                tm,x=lis
                cdef np.ndarray[DTYPE_t,ndim=1] a=np.zeros((18),dtype=DTYPE)
                return a

#User supplied function dy/dt = f. We can use the np array and related operations to define it, including array reshape and concatenate, which simplifys the process.
cdef class userSupply(Function):
        cpdef np.ndarray[DTYPE_t,ndim=1] evaluate(self,double t,np.ndarray[DTYPE_t,ndim=1] y,np.ndarray[DTYPE_t,ndim=1] lis):
                cdef tm,x
                tm,x=lis
                j1 = EXP(-32.*(t-9./16)**2)
                j2 = EXP(-32.*(t-7./16)**2)
                cdef np.ndarray[DTYPE_t,ndim=2] H=np.zeros((3,3),dtype=DTYPE)
                H[0,1]=j1;H[1,0]=j1;H[1,2]=j2;H[2,1]=j2
                rhoR=y[0:9].reshape((3,3))
                rhoC=y[9:18].reshape((3,3))
                dRdt=np.dot(H,rhoC)-np.dot(rhoC,H)-x*(rhoR-np.diag(rhoR))
                dCdt=np.dot(rhoR,H)-np.dot(H,rhoR)-x*(rhoC-np.diag(rhoC))
                return np.pi*tm*np.concatenate((dRdt.reshape((9)),dCdt.reshape((9))))

#compute_dephase impliments the loop within which the equation is solved with different parameters each time.
cdef compute_dephase(Function f,double N1,double N2):
        output1=open('dataDephase1','w')
        output2=open('dataDephase2','w')
        output3=open('dataDephase3','w')
        cdef int dimension=18
        cdef double x,tm
        cdef double t,h
        cdef np.ndarray[DTYPE_t,ndim=1] arg=np.zeros((2),dtype=DTYPE)
        cdef np.ndarray[DTYPE_t,ndim=1] y=np.zeros((dimension),dtype=DTYPE)
        for x in np.linspace(0.,0.5,N1):
            for tm in np.linspace(0,10,N2):
                arg[0]=tm;arg[1]=x
                step = odeiv.step_rkf45(dimension,f.evaluate,args=arg)
                control = odeiv.control_yp_new(step,1e-7,1e-7)
                evolve = odeiv.evolve(step, control,dimension)
                y[0]=1.0
                t=0.;h = 1.e-4
                while t<1.:
                    if (1.-t)<h: h=1.-t
                    t, h, y = evolve.apply(t,1.,h,y)
                output1.write(" %f " % y[0])
                output2.write(" %f " % y[4])
                output3.write(" %f " % y[8])
                y=np.zeros((dimension),dtype=DTYPE)
            output1.write("\n")
            output2.write("\n")
            output3.write("\n")
        output1.close()
        output2.close()
        output3.close()

cpdef callFunction(Function f,double N1,double N2):
        compute_dephase(f,N1,N2)
\end{lstlisting}

\lstset{language=Python,caption={The result of
profiling the Cython code and the Python one.},label=profileDephase}
\begin{lstlisting}
#Profile result for dephaseCython.pyx. 
#The CPU of my laptop is 2 Intel(R) Core(TM)2 Duo CPU T7500 @2.20GHz.
Tue May  3 21:24:15 2011    Profile.prof

         314988757 function calls in 4075.919 seconds

   Ordered by: internal time

   ncalls  tottime  percall  cumtime  percall filename:lineno(function)
  3987941 2750.219    0.001 4032.149    0.001 {pygsl.__callback.gsl_odeiv_evolve_apply}
 61192534  693.108    0.000 1281.930    0.000 twodim_base.py:220(diag)
 61192534  229.666    0.000  413.172    0.000 numeric.py:216(asarray)
 61192534  183.506    0.000  183.506    0.000 {numpy.core.multiarray.array}
122385068  175.650    0.000  175.650    0.000 {len}
        1   23.031   23.031 4075.919 4075.919 {dephaseCython.callFunction}
  3987941   17.043    0.000 4049.192    0.001 odeiv.py:421(apply)
    50000    0.860    0.000    1.364    0.000 odeiv.py:385(__init__)
    50000    0.392    0.000    0.649    0.000 odeiv.py:88(__init__)
   200000    0.318    0.000    0.318    0.000 {hasattr}
    50000    0.287    0.000    0.476    0.000 odeiv.py:407(__del__)
    50000    0.273    0.000    0.439    0.000 odeiv.py:260(__del__)
    50000    0.263    0.000    0.447    0.000 odeiv.py:110(__del__)
    50000    0.195    0.000    0.317    0.000 odeiv.py:365(__init__)
    50000    0.166    0.000    0.166    0.000 {pygsl.__callback.gsl_odeiv_step_alloc}
    50000    0.122    0.000    0.122    0.000 {pygsl.__callback.gsl_odeiv_control_yp_new}
    50000    0.117    0.000    0.117    0.000 {pygsl.__callback.gsl_odeiv_evolve_alloc}
    50000    0.110    0.000    0.110    0.000 {pygsl.__callback.gsl_odeiv_step_free}
    50000    0.110    0.000    0.110    0.000 {pygsl.__callback.gsl_odeiv_evolve_free}
    50000    0.091    0.000    0.091    0.000 {pygsl.__callback.gsl_odeiv_control_free}
    50000    0.080    0.000    0.080    0.000 odeiv.py:163(_get_func)
    50000    0.077    0.000    0.077    0.000 odeiv.py:166(_get_jac)
    50000    0.077    0.000    0.077    0.000 odeiv.py:169(_get_args)
    50000    0.077    0.000    0.077    0.000 odeiv.py:160(_get_ptr)
    50000    0.076    0.000    0.076    0.000 odeiv.py:293(_get_ptr)
      101    0.003    0.000    0.004    0.000 function_base.py:6(linspace)
      101    0.001    0.000    0.001    0.000 {numpy.core.multiarray.arange}
        1    0.000    0.000 4075.919 4075.919 <string>:1(<module>)
        1    0.000    0.000    0.000    0.000 {method 'disable' of '_lsprof.Profiler' objects}


#Profile result for dephase.py.
Tue May  3 23:33:12 2011    Profile.prof

         651748000 function calls in 6163.858 seconds

   Ordered by: internal time

   ncalls  tottime  percall  cumtime  percall filename:lineno(function)
 30596267 2708.150    0.000 5925.743    0.000 dephase.py:14(evaluate)
 91788801 1223.705    0.000 1223.705    0.000 {numpy.core.multiarray.array}
 61192534  674.102    0.000 1259.449    0.000 twodim_base.py:220(diag)
122385068  455.155    0.000  455.155    0.000 {numpy.core.multiarray.dot}
122385068  363.905    0.000  363.905    0.000 {method 'reshape' of 'numpy.ndarray' objects}
 61192534  227.731    0.000  411.719    0.000 numeric.py:216(asarray)
  3987941  201.941    0.000 6127.684    0.002 {pygsl.__callback.gsl_odeiv_evolve_apply}
122385068  173.628    0.000  173.628    0.000 {len}
 30596267   99.369    0.000   99.369    0.000 {numpy.core.multiarray.concatenate}
        1   16.182   16.182 6163.858 6163.858 dephase.py:26(compute_dephase)
  3987941   15.966    0.000 6143.650    0.002 odeiv.py:421(apply)
    50000    0.805    0.000    1.295    0.000 odeiv.py:385(__init__)
    50000    0.357    0.000    0.597    0.000 odeiv.py:88(__init__)
   200000    0.308    0.000    0.308    0.000 {hasattr}
   150300    0.301    0.000    0.301    0.000 {method 'write' of 'file' objects}
    50000    0.278    0.000    0.457    0.000 odeiv.py:407(__del__)
    50000    0.263    0.000    0.425    0.000 odeiv.py:260(__del__)
    50000    0.255    0.000    0.434    0.000 odeiv.py:110(__del__)
    50000    0.214    0.000    0.214    0.000 {numpy.core.multiarray.zeros}
    50000    0.186    0.000    0.299    0.000 odeiv.py:365(__init__)
    50000    0.158    0.000    0.158    0.000 {pygsl.__callback.gsl_odeiv_step_alloc}
    50000    0.113    0.000    0.113    0.000 {pygsl.__callback.gsl_odeiv_control_yp_new}
    50000    0.109    0.000    0.109    0.000 {pygsl.__callback.gsl_odeiv_evolve_alloc}
    50000    0.105    0.000    0.105    0.000 {pygsl.__callback.gsl_odeiv_step_free}
    50000    0.102    0.000    0.102    0.000 {pygsl.__callback.gsl_odeiv_evolve_free}
    50000    0.088    0.000    0.088    0.000 {pygsl.__callback.gsl_odeiv_control_free}
    50000    0.078    0.000    0.078    0.000 odeiv.py:163(_get_func)
    50000    0.077    0.000    0.077    0.000 odeiv.py:166(_get_jac)
    50000    0.076    0.000    0.076    0.000 odeiv.py:293(_get_ptr)
    50000    0.075    0.000    0.075    0.000 odeiv.py:169(_get_args)
    50000    0.075    0.000    0.075    0.000 odeiv.py:160(_get_ptr)
      101    0.003    0.000    0.003    0.000 function_base.py:6(linspace)
      101    0.001    0.000    0.001    0.000 {numpy.core.multiarray.arange}
        3    0.000    0.000    0.000    0.000 {open}
        3    0.000    0.000    0.000    0.000 {method 'close' of 'file' objects}
        1    0.000    0.000 6163.858 6163.858 <string>:1(<module>)
        1    0.000    0.000    0.000    0.000 {method 'disable' of '_lsprof.Profiler' objects}


\end{lstlisting}

\section{Conclusion}

The above examples are about how to solve the ODE problems in Python.
One can also do monte carlo simulation in Python. There are many
packages that can be used in Python, such as ALPS (Algorithms and
Libraries for Physics Simulations)~\cite{alps1,alps2} and
PyMC~\cite{pymc}.  GSL also has modules for Monte Carlo simulation  ,
which can be used via Pygsl as above.  Python is powerful and easy to
learn. Its syntax is simple and most of the time, it is where to find
the libraries needed and combine them together that provides difficult
for beginners. Once get familiar with it, it would be  elegant and
        intuitive to do numerical simulations. We hope this can help
        non-computation specialists get familiar with Python and
        implement their own models efficiently in it.

\begin{acknowledgments}
Project supported by the National Natural Science Foundation of China 
(Grant No.\ 10847150), the Natural Science Foundation of Shandong 
Province (Grant No.\ ZR2009AM026), Scientific Research Foundation 
for Returned Scholars, Ministry of Education of China, and Research 
Project of  Key Laboratory for Magnetism and Magnetic Materials 
of the Ministry of Education, Lanzhou University.
\end{acknowledgments}

\bibliography{python}

\end{document}